\documentclass[preprint,showpacs,preprintnumbers,amsmath,amssymb]{revtex4}%
\usepackage{graphicx}
\usepackage{dcolumn}
\usepackage{bm}
\usepackage{amsmath}
\usepackage{amsfonts}
\usepackage{amssymb}%
\usepackage{appendix}
\providecommand{\U}[1]{\protect\rule{.1in}{.1in}}
\newcommand{\be}{\begin{equation}}
\newcommand{\en}{\end{equation}}
\newcommand{\bea}{\begin{eqnarray}}
\newcommand{\ena}{\end{eqnarray}}
\begin{document}
\title{Reconstructing G-inflation: From the attractors $n_S(N)$ and $r(N)$  }
\author{Ram\'on Herrera}
\email{ramon.herrera@pucv.cl} \affiliation{Instituto de
F\'{\i}sica, Pontificia Universidad Cat\'{o}lica de
Valpara\'{\i}so, Avenida Brasil 2950, Casilla 4059,
Valpara\'{\i}so, Chile.}

\date{\today}

\begin{abstract}

The reconstruction of an inflationary universe in the context of the
 Galileon model or G-model, considering as attractors the
scalar spectral index   $n_S(N)$ and the tensor to scalar ratio
$r(N)$ as a function of the number of e-folding  $N$ is studied.
By assuming a coupling of the form $G(\phi,X)=g(\phi)\,X$, we obtain
the effective potential $V$ and the
coupling parameter $g$ in terms of the cosmological parameters
$n_S$ and $r$
under the slow roll approximation.  From some examples for $n_S(N)$ and $r(N)$, different results for the
effective potential $V(\phi)$ and the coupling parameter
$g(\phi)$ are found.

\end{abstract}

\maketitle

\section{Introduction}

It is well known that during the dynamic evolution of the early
universe, it presented a period of  rapid  growth   called
inflationary stage or merely inflation
 \cite{R1,R102,R103,Rin}.
 The  inflationary  universe
gives an elegant solutions to long standing cosmological problems
present in the standard hot big bang model. Nevertheless, inflation not only solves
the problems of the hot big bang, but also
 gives account of    the Large-Scale
Structure (LSS) \cite{R2,R203}, together with a causal description
of the anisotropies observed in the Cosmic Microwave Background
(CMB) radiation of the early
universe\cite{astro,Hinshaw:2012aka,Ade:2013zuv,Planck2015,Ob2,DiValentino:2016foa}.

In the context of the different models that give account of   the
dynamical evolution  of an inflationary universe, we can stand out
 one  general class of inflationary models
 where  the inflation
 is driven by a minimally coupled scalar field.
 In the literature these models are called
the Galilean  inflationary models  or simply G-inflation and
its generalization, known  as G$^2-$inflation
which  also corresponds to a subclass of the
 Horndeski theory\cite{Ho},
was developed  in Ref.\cite{Kobayashi:2011nu}.
In this
context, the inclusion of the canonical and non-canonical scalar
field in the model of G-inflation, is known as kinetic gravity
braiding model\cite{G1,G2}.  From the
observational point the view, the detection of  gravitational waves by
GW170817 and  the $\gamma-$ray burst
\cite{TheLIGOScientific:2017qsa,Monitor:2017mdv,GBM:2017lvd} give
  a strong constrain on the speed gravitational  waves. In this context, the
  G-model  is consistent with the GW170817, since the speed gravitational  waves
  is equal to the speed of
light. Thus,
  the Galilean  action  includes an extra  term of the form
 $G(\phi,X)\square\phi$ to the standard action, but this term on the action  does not modify
the speed gravitational  waves and it is equivalent to the speed of
light
 \cite{G1,G2}.  We should also mention that
 the field equations still have
derivatives only up to second order, see Ref.\cite{Nic}. In this way,  different
cosmological  models have been developed  in the framework
of G-inflation. In particular,
 assuming the  slow roll approximation and considering
 some effective potentials G-inflation was studied in Ref.\cite{288}.
In relation to the Higgs field, the model of Higgs G-inflation
viewed as a modification of the standard Higgs inflation, in which
the function $G(\phi,X)\propto \phi X$,   was developed in
Ref.\cite{289a}, see also\cite{289}. In this modified gravity, the specific case  in which the effective
potential $V(\phi)=$ constant together with the slow roll approximation,  is  known
 as  ultra slow roll G-inflation and it  was
studied in \cite{Ut}. For the case in which the scalar potential
is  of the  power-law type was studied in Ref.\cite{Pw}. The model  of
warm inflation and its thermal fluctuations in the context of
G-inflation  was developed in Ref.\cite{Herrera:2017qux}. The reheating mechanism in
this model was studied in  Ref.\cite{Br}
  and from string gas cosmology in Ref.\cite{agr}, see also\cite{Reh2,DE1,DE2,Ka}.


On the other hand, the idea of the reconstruction of the
physical   variables present  on the background dynamics of
  inflationary models, from
observational parameters such as the scalar spectrum,  scalar
spectral index $n_S$ and the tensor to scalar ratio $r$,   have
been studied  by several
authors\cite{H1,H2,H3,H4,M,Chiba:2015zpa,H5}.  In this context, a
reconstruction mechanism to obtain the physical variables in the
inflationary stage  considering the slow roll approximation,  is though of
 the parametrization of the cosmological observables   $n_S(N)$ and $r(N)$
  or commonly called attractors, in which the parameter $N$ denotes  the
number of e-folds.

From the observational point of view, the
scalar spectral index  $n_S$,
 is well supported by the parametrization in terms of  the
number of e-fonds $N$, given by the attractor   $n_S\sim 1-2/N$ for large $N$,
in which the number $N\simeq 50-70$ at the end of the
inflationary scenario from the results of Planck and BICEP2-Keck Array Collaborations
\cite{Planck2015,Ob2}.
 In the framework of the General Relativity (GR), different models can be
reconstructed  considering the parametrization or attractor $n_S(N)\sim 1-2/N$
(for large $N\sim\,\mathcal{O}(10^{2}) $),
to name of few we have; the hyperbolic tangent model or simply the
T-model \cite{T}, E-model\cite{E},
$R^2$-model\cite{R102}, the chaotic inflationary model\cite{R103}, the
 Higgs inflation \cite{Higgs,Higgs2}, etc. In the context of warm inflation and
 its reconstruction  was necessary to introduce
 the attractors $n_S(N)$ and $r(N)$ (unlike cold inflation),
 in order to build  the
 scalar potential and the dissipation coefficient  in terms of the scalar field \cite{Herrera:2018cgi}.

Another methodology used  in the literature in
order  to reconstruct   the scalar potential, scalar spectral index
and the tensor to scalar ratio is by means
of the slow-roll
parameter $\epsilon(N)$, as function of the  number of e-folds $N$
\cite{Huang:2007qz,M,Gao:2017owg,M}. Also, the use
 of   two slow
roll parameters $\epsilon(N)$ and $\eta(N)$,  for the reconstruction
of the scalar potential and spectral index was assumed  in Ref.\cite{Roest:2013fha}, see also Refs.\cite{N1,
N2}.


The goal of this investigation is to reconstruct the Galilean inflationary  model or
G-inflation, given the parametrization of
 the scalar spectral index and the
tensor to scalar ratio,  in terms of the number of e-folds.
In this
sense, we investigate  how the Galilean inflationary  model in which the function $G(\phi,X)$,
is given by
$G(\phi,X)=g(\phi)\,X$,
 modifies  the
reconstructions  of the scalar potential $V(\phi)$ and
 the coupling parameter $g(\phi)$, from the attractor point
$n_S(N)$ and $r(N)$.    Thus, we will determine  the structure of
the function $g(\phi)$ and in particular of $V(\phi)$, in order to
in   account of the observables  $n_S$ and the ratio $r$ given by the observations.

By considering the domination of the Galilean effect, we develop
  a general formalism in order to obtain the effective potential $V$ and coupling
  parameter $g$,
   from the parametrization of the cosmological
attractors $n_S(N)$ and $r(N)$, under the slow roll approximation.

For the application of the   developed formalism, we will study different  examples for the attractor
 point $n_S(N)-r(N)$.  From these attractors, we will reconstruct the
 effective  potential $V(\phi)$ and the coupling parameter $g(\phi)$ in terms of the scalar field $\phi$. Also, we
 will obtain
 different
 constraints on the  parameters present in the reconstruction.


The outline of the paper is as follows: The next section we give
 a brief description  of the model of  G-inflation. Here, the background
 dynamics and cosmological perturbations are presented.
 In the section III, we develop a general formalism  in order to reconstruct the scalar
 potential and coupling parameter in function of the attractors $n_S(N)$
 and $r(N)$, respectively.
   In section IV we
apply the method for different   examples of $n_S(N)$
 and $r(N)$ so as  to construct the effective potential $V(\phi)$ and the coupling $g(\phi)$
 in terms of the scalar field $\phi$.
 Finally, in section V  we give  our conclusions.
 We chose units so that
$c=\hbar=8\pi=1$.

\section{ The model of G-inflation}
In this section we give a brief description of the model of G-inflation. We start with
 the 4-dimensional action for the Galilean model  given by
\begin{equation}
S = \int \sqrt{-g_{4}}d^{4}x\,\left(\frac{M^{2}_{P}}{2}R
+K(\phi,X)-G(\phi,X)\square\phi \right),\label{action}
\end{equation}
where  $g_{4}$ denotes the determinant of the space-time metric
$g_{\mu\nu}$,  $R$ is the Ricci scalar and the quantity $X$
corresponds to
$X=-g^{\mu\nu}\partial_{\mu}\phi\partial_{\nu}\phi/2$, where
$\phi$ denotes the scalar field. The quantities   $K$ and $G$ are
arbitrary functions of $X$ and the scalar field $\phi$, respectively. Here,  the
quantity  $M_p$ corresponds to  the Planck mass.

By assuming a spatially  flat  Friedmann Robertson Walker
(FRW) metric, along with  a scalar field homogeneous in which
$\phi=\phi(t)$, the    Friedmann equation can be written as
\begin{equation}
3H^{2} = \kappa\,\rho .\label{HC}
\end{equation}
Here, the parameter $H=\frac{\dot{a}}{a}$ denotes the Hubble rate,
$a$ corresponds to the scale  factor and $\rho$ is the energy density.  In the following,  the dots denote
differentiation with  respect to the time and  the quantity  $\kappa=1/M_p^2$.

 From the action (\ref{action}),  we can identify that the energy
density and the pressure associated to the scalar field $\phi$ are
given by
\cite{G1,G2}

\begin{equation}
\rho = 2K_X\,X-K+3G_XH\dot{\phi}^3-2G_\phi X,\label{r1}
\end{equation}
and
\begin{eqnarray}
p = K-2(G_\phi+G_X\ddot{\phi})X,\label{pp1}
\end{eqnarray}
respectively. In the following, we will assume  that the notation  $K_X$
denotes
 $K_X=\partial K/\partial X$, $G_\phi$ corresponds to $G_\phi=\partial G/\partial\phi$,
 $K_{XX}=\partial ^2K/\partial X^2$,  etc.

In this framework, the continuity  equation for the energy density
$\rho$ can be written as $ \dot{\rho}+3\,H\,(\rho+p)=0,$ or
equivalently
$$
  K_X\square\phi+2K_{XX}X\ddot{\phi}+2K_{X\phi}X-K_{\phi}-2(G_\phi-G_{X\phi}X)\square\phi
  +6G_X(\dot{H}X+\dot{X}H
  +3H^2X)
  $$
  \begin{equation}
 +6HG_{XX}X\dot{X} -2G_{\phi\phi}X-4G_{X\phi}X\ddot{\phi}=0,
 \;\;\;\;\mbox{where}\,\,\,\,\;\;\square\phi=\ddot{\phi}+3H\dot{\phi}.\label{eq7}
\end{equation}
Here, we have used
 Eqs.(\ref{r1}),  and  (\ref{pp1}).

In particular, for the special  cases  in which $K=X-V(\phi)$ and
$G=0$,  in which $V(\phi)$ corresponds
to the effective potential, we recovered the standard General Relativity (GR).

In order to study the reconstruction for the G-model,
 we will analyze  the special case  in which  the functions $K(\phi,X)$ and $G(\phi,X)$
are given by
 \begin{equation}
   K(\phi,X)=X-V(\phi),\;\;\;\;\mbox{and}\;\;\;G(\phi,X)=-g(\phi)\,X,\label{anz}
 \end{equation}
respectively. Here, $g(\phi)$ is a function that depends exclusively on the scalar
field $\phi$.

  From Refs.\cite{G1,G2}, we will assume the slow roll approximation. In this
  context,
the   potential $V(\phi)$ dominates over the quantities $X$, $|G_X
H\dot{\phi}^3|$ and $|G_\phi X|$, wherewith the energy density
  $\rho_{\phi}\sim V(\phi)$ and then the
   Friedmann equation (\ref{HC}) becomes
\begin{equation}
3H^{2}\approx\,\kappa\,
V(\phi).\label{HH}
\end{equation}

Introducing  the slow-roll parameters $\epsilon_1$, $\epsilon_2$, $\epsilon_3$  and
$\epsilon_4$ given by  \cite{G2}
\begin{equation}
\epsilon_1=-\frac{\dot{H}}{H^2},\,\,\,\epsilon_2=-\frac{\ddot{\phi}}{H\dot{\phi}},
\,\,\,\epsilon_3=\frac{g_\phi\dot{\phi}}{gH},\,\,\mbox{and}\,\,\,\epsilon_4=\frac{g_{\phi\phi}X^2}{V_{\phi}},\label{pr}
\end{equation}
then  the Eq.(\ref{eq7}) can be rewritten as

\begin{equation}
3H\dot{\phi}(1-\epsilon_2/3-gH\dot{\phi}[3-\epsilon_1-2\epsilon_2-2\epsilon_2\epsilon_3/3])=
-(1-2\epsilon_4)V_\phi.\label{eqf}
\end{equation}
Here, we have used  the functions given by Eq.(\ref{anz}) and the slow roll parameters given by
Eq.(\ref{pr}).

By considering that the slow-roll parameters $\epsilon_1$,
$|\epsilon_2|$, $|\epsilon_3|$, $|\epsilon_4|\ll 1$, then the
Eq.(\ref{eqf}) is reduced to
\begin{equation}
3H\dot{\phi}(1+{\cal{A}})\simeq
-V_\phi,\;\;\;\;\mbox{where}\;\;\;{\cal{A}}=-3gH\dot{\phi}.
\label{scalar1}
\end{equation}

Here, we mention that in relation to the slow roll equation
(\ref{scalar1}), we have two limiting instances. The situation in
which $|{\cal{A}}|\ll 1$ corresponds to the standard equations of GR in
the framework of slow roll inflation. Instead, the inverse case in
which $ |{\cal{A}}|\gg 1$, the Galileon effect changes the dynamic
equation of the scalar field $\phi$ and hence the dynamics of
inflationary model. In this context, we are interested in the latter situation
in which  the Galileon effect modifies the  dynamics of the G-model and its reconstruction. Thus,
$3H\dot{\phi}{\cal{A}}\simeq -V_\phi$ and then
$9H^2\dot{\phi}^2\simeq (V_\phi/g) $ suggesting  that the ratio
$(V_\phi/g)>0$. Therefore, in the case in which $V_\phi>0$ then
the quantity $g>0$ and vice versa. In the following we shall take
$V_\phi>0$ and $g > 0$.

 Typically, if the scalar field roll down
potential, then the velocity of the scalar field can be written as
\begin{equation}
\dot{\phi}\simeq -\sqrt{\frac{V_\phi}{3\kappa\, g\,V}}.
\end{equation}
Here, we have considered Eq.(\ref{HH}). Also, we note that the
parameter ${\cal{A}}>0$, since we have assumed that $\dot{\phi}<0$.

In relation to the expansion, we define the number of e-folding
$N$ in order to give
 a measure of the inflationary expansion. In this way, we assume     two
different values of cosmological times $t$ and $t_e$. Here,
 the
time $t_e$ denotes  the end of inflationary epoch. Thus, the
number of e-folds $N$ in the limit in which $1\ll |{\cal{A}}|$
is given by
\begin{equation}
N=\int_t^{t_e}H\,dt'=\int_\phi^{\phi_e}H\,\frac{d\phi'}{\dot{\phi}}\simeq
\kappa\,\int_{\phi_e}^{\phi}\,V\,\left[\frac{g}{V_{\phi'}}\right]^{1/2}\,d\phi'.\label{3N}
\end{equation}

On the other hand,  the cosmological perturbations together with
the scalar and tensor spectrums were obtained in
Refs.\cite{Kobayashi:2011nu,GG1,GG2,Ohashi:2012wf} for the model
of G-inflation. In this sense, from the action (\ref{action}), the
amplitude of scalar perturbations ${\cal{P_S}}$ generated during
the inflationary epoch for a flat space and assuming the slow roll
approximation  we have

\begin{equation}
{\cal{P_S}}\simeq\frac{\kappa^3\,V^3}{12\pi^2\,V_\phi^2}\,\frac{(1+{\cal{A}})^2
(1+2{\cal{A}})^{1/2}}{(1+4{\cal{A}}/3)^{3/2}}.\label{P1}
\end{equation}

Because the scalar spectral index $n_S$ is defined as
$n_S=d\ln{\cal{P_S}}/d\ln k$, then from Eq.(\ref{P1}) the index
$n_S$  in terms of the standard slow roll parameters can be
written as

\begin{equation}
n_S-1\simeq\,-\frac{6\epsilon}{1+{\cal{A}}}+\frac{2\eta}{1+4{\cal{A}}/3}
\left[1-\frac{{\cal{A}}}{6(1+2{\cal{A}})^2}\right],\label{ns0}
\end{equation}
where the standard parameters $\epsilon$ and $\eta$ are defined as
\begin{equation}
  \epsilon=\frac{1}{2\kappa}\left(\frac{V_{\phi}}{V}\right)^2,\;\,\,
\,\,\,\mbox{and}\,\,\,\eta=\frac{V_{\phi\phi}}{\kappa\,V}.\label{CTM}
\end{equation}

Note that in the limit $g\rightarrow 0$ (or equivalently
${\cal{A}}\rightarrow 0$), the spectral index given by
Eq.(\ref{ns0}) coincides with the GR in which $n_S-1\simeq
-6\epsilon+2\eta$.
By assuming the limit in which $|{\cal{A}}|\gg 1$, the scalar spectral index reduces
to
\begin{equation}
n_S-1\simeq-\frac{6\epsilon}{{\cal{A}}}+
\frac{3\eta}{2\cal{A}}.\label{ns}
\end{equation}

On the other hand,  in relation to  the tensorial perturbations, the amplitude
 of the tensor mode was determined in refs.\cite{Kobayashi:2011nu,GG1,GG2,Ohashi:2012wf}, and
 then
  the tensor spectrum ${\cal{P_T}}$ is defined as
${\cal{P_T}}\simeq (2\kappa^2V/3\pi^2)$. In this form, the tensor
to scalar ratio $r$
 in the
framework of  G-inflation can be written as

\begin{equation}
r=\frac{{\cal{P_T}}}{{\cal{P_S}}}\simeq\,16\epsilon\,\left[\frac{(1+4{\cal{A}}/3)^{3/2}}
{(1+{\cal{A}})^2(1+2{\cal{A}})^{1/2}}\right].\label{rl}
\end{equation}
One again, note that in the limit ${\cal{A}}\rightarrow 0$ ( or
equivalently $g\rightarrow 0$), the ratio $r$ coincides with that
corresponding to GR in which  $r=16\epsilon$.

Taking the limit $ |{\cal{A}}|\gg 1$, the tensor to scalar ratio $r$ results
\begin{equation}
r\simeq\frac{4\sqrt{2}}{3^{3/2}}\,\frac{16\epsilon}{\cal{A}}.\label{r}
\end{equation}

In the following we will analyze the reconstruction of the effective
potential $V(\phi)$ and the  coefficient $g(\phi)$ in the
framework
 of
G- inflation. In order to realize the reconstruction we will assume
 the limit $1\ll |{\cal{A}}|$, together with
 an attractor point from the   index $n_S(N)$ and
the ratio $r(N)$ on the $r-n_S$ plane.

\section{Reconstructing  G-inflation \label{secti2a} }

In this section we develop the method to follow in order to
reconstruct, the scalar potential $V(\phi)$ and the coupling  parameter
$g(\phi)$, assuming the scalar spectral index $n_S(N)$ and the
tensor to scalar ratio $r(N)$ as attractors. In order to
reconstruct analytically  the potential $V$ and the coupling
parameter $g$, we shall take the limit  $|{\cal{A}}|\gg 1$.
 Following Refs.\cite{Chiba:2015zpa,Herrera:2018cgi},
we rewrite    the  spectral index and the tensor to
scalar ratio given by Eqs.(\ref{ns}) and (\ref{r}), in terms of the number of e-folds $N$ and its derivatives.
 Thus, from these
relations  and giving  $n_S=n_S(N)$ and $r=r(N)$,  we should find
the  potential $V$ and the coupling  parameter $g$ as a function
of the number $N$. Later, with the help of Eq.(\ref{3N}) we should
obtain the e-folding $N$ in terms of the scalar field $\phi $ in
order to reconstruct finally, the scalar potential $V(\phi)$ and the coupling
parameter $g(\phi)$, respectively.

 In this sense, we begin  by rewriting
 the standard  slow roll parameters    as a function  of the number of e-folds,
taking into account that
$$
V_{\phi}=\frac{dV}{d\phi}=\kappa V\sqrt{\frac{g}{V_{\phi}}}
\,V_{\,N}\,,
$$
then we get
\begin{equation}
V_{\phi}=(\kappa V\sqrt{g}\,\,V_{\,N})^{2/3}\,,\,\,\,\,\mbox{wherewith}\,\,\,\,V_{\,N}=\frac{dV}{dN}>0.\label{dV}
\end{equation}
In the following, we will assume  the subscript $V_{\,N}=dV/dN$,
$V_{NN}$ to $V_{NN}=d^2V/dN^2$, $g_{_{N}}=dg/dN$ etc.

Similarly  for $V_{\phi\phi}$ we get
\begin{equation}
V_{\phi\phi}=\frac{2}{3}\,(\kappa^4\,V^4\,g^2\,V_{N})^{1/3}\,\left[\frac{V_{N}}{V}+
\frac{ (\sqrt{g})_{N}}{\sqrt{g}}+\frac{V_{\,NN}}{V_{\,N}}\right].\label{ddV}
\end{equation}

In this form, the standard slow roll parameters $\epsilon$ and
$\eta$ are rewritten as
\begin{equation}
\epsilon=\frac{1}{2\kappa V^2}\,(\kappa V \sqrt{g}\,V_{\,N})^{4/3},\label{p1}
\end{equation}
and
\begin{equation}
\eta=\frac{2}{3\kappa V}\,(\kappa^4\,V^4\,g^2\,V_{\,N})^{1/3}\,\left[\frac{V_{\,N}}{V}+
\frac{ (\sqrt{g})_{N}}{\sqrt{g}}+\frac{V_{\,NN}}{V_{\,N}}\right]
,\label{p2}
\end{equation}
respectively.

Now, the relationship  between the e-folds $N$ and the scalar
field $\phi$, from Eq.(\ref{3N}) can be written as
 \begin{equation}
   \int
   \left[\frac{V_{\,N}}{\kappa^2\,g\,V^2}\right]^{1/3}\,dN=\int\,d\phi.\label{NF}
 \end{equation}

In this context, by considering   Eqs.(\ref{ns}),
(\ref{p1}) and (\ref{p2}) we find that  the scalar spectral index $n_S$ can be rewritten in
terms of the e-folding  $N$,  such that
$$
n_S-1=-2\frac{V_{\,N}}{V}+\left[\frac{
(\sqrt{g})_{N}}{\sqrt{g}}+\frac{V_{\,NN}}{V_{\,N}}\right],
$$
or equivalently
\begin{equation}
n_S-1=\left[\ln \left(\frac{\sqrt{g}\,\,V_{\,N}}{V^2}\right)\right]_{\,N}.\label{ns1}
\end{equation}
From Eq.(\ref{r}) we obtain that   the tensor to scalar ratio becomes
\begin{equation}
  r=\frac{{\cal{P_T}}}{{\cal{P_S}}}=C\,\frac{V_{\,N}}{V},\,\,\,\,\mbox{in
  which}\,\,\,\;\;
  \,C=\frac{32\sqrt{6}}{9}\simeq\,8.71.\label{gr8}
\end{equation}
 Here, we have considered that the function $\cal{A}$ can be rewritten as
 $
 {\cal{A}}=-3gH\dot{\phi}=\sqrt{g}\,(\kappa\,V\,\sqrt{g}\,V_{\,N})^{1/3}.
 $
Also, we note that in the context of the reconstruction and
assuming ${\cal A}\gg 1$, the ratio $r$ given by Eq.(\ref{gr8})
does not depend of the parameter $g$. Here, we noted that  this ratio $r$
 is similar to the one obtained in the standard  GR, where $r=8V_{N}/V$, see Ref.\cite{Chiba:2015zpa}.

Now, from Eqs.(\ref{ns1}) and (\ref{gr8})
we obtain that the effective potential in terms of the e-folding $N$
results
\begin{equation}
V(N)=\,\exp\left[\frac{1}{C}\int\,r\,dN\right],\label{Vw}
\end{equation}
 and the coupling parameter
$g(N)$ becomes
\begin{equation}
g(N)=\frac{C^2}{r^2}\,\exp\left[2\int\,\left(\frac{r}{C}+[n_S-1]\right)dN\right]
.\label{GGd}
\end{equation}
Here, we emphasize  that the potential $V(N)$ depends only of the tensor to scalar ratio $r(N)$.

In fact,  the Eqs.(\ref{NF}), (\ref{Vw}) and
(\ref{GGd}) are the fundamental   equations in order to reconstruct
of the effective potential $V(\phi)$ and the
parameter $g(\phi)$,   giving the attractors
$n_S(N)$ and $r(N)$, respectively.

In the following we will study some specific examples in order to reconstruct
the scalar potential $V(\phi)$ and coupling parameter $g(\phi)$, from the
cosmological parameters or attractors $n_S(N)$ and $r(N)$, respectively.

\section{Some examples of reconstruction}

In order to  apply the formalism of above, we shall first consider the simplest
example for the attractors   $n_S(N)$ and $r(N)$, so as
 to reconstruct analytically the effective potential
$V(\phi)$ and coupling  parameter $g(\phi)$. In this sense and
 following
Refs.\cite{T,Chiba:2015zpa}, we assume that the spectral
index in terms of the number of e-folds  is given by
\begin{equation}
n_S-1=-\frac{2}{N},\label{e1}
\end{equation}
and the tensor to scalar ratio as
\begin{equation}
r=\frac{\alpha}{N},\label{e2}
\end{equation}
where $\alpha>0$ corresponds to a dimensionless  constant. We mention that in
 the framework of GR, for the chaotic model (in which $V(\phi)\propto \phi^2$)
 \cite{R103},
 the parametrization in terms of $N$ of
  the scalar index is given by
Eq.(\ref{e1}), where the value of the parameter $\alpha=8$,
i.e., $r(N)=8/N$, see Ref.\cite{Chiba:2015zpa}.
 In particular
if we consider that the number $N$ before the end of inflationary
scenario occurs at
  $N \simeq 60$, then the tensor to scalar ratio given by Eq.(\ref{e2})
  is well corroborated by observational
 data when the constant $\alpha<4.2$. Here, we have used that  the ratio
  $r<0.07$ from BICEP2 and Keck Array Collaborations\cite{Ob2}.

From the attractor given by Eq.(\ref{e2}), we find that the
effective potential (\ref{Vw}) results
\begin{equation}
V(N)=V_0\,N^{\frac{\alpha}{C}},\label{pot1}
\end{equation}
where $V_0>0$ corresponds to the integration constant (with units
of $M_p^4$). By utilizing the spectral index  (\ref{e1}), we
obtain that the integral expression
$\exp[\int(n_s-1)]dN=\beta/N^2$, in which $\beta$ denotes an
integration constant (with units of $M_p^{-11/2}$ or
$\kappa^{11/4}$). In this form, we obtain that the coupling
parameter $g(N)$ in terms of e-folds from Eq.(\ref{GGd}) can be
written as
\begin{equation}
g(N)=
g_0\,N^{\frac{-2(C-\alpha)}{C}},\,\,\,\,\,\mbox{where}\,\,\,\,\,g_0=\left(\frac{\beta
C V_0}{\alpha}\right)^2.\label{g1}
\end{equation}
By using Eqs.(\ref{pot1}) and (\ref{g1}), we find that the
parameter $\cal{A}$ in terms of the number of e-folds results
\begin{equation}
{\cal{A}}(N)={\cal{A}}_0\,N^{\frac{-(5C-6\alpha)}{3C}}\,,\;\;\;\;\mbox{in
which}\;\;\;\;\;
{\cal{A}}_0=\left[\frac{C\,V_0^2\,\kappa^{1/3}\beta^{4/3}}{\alpha}\right].\label{AA}
\end{equation}
From the condition in which predominate the Galileon effect, such
that ${\cal{A}}\gg 1$, then from Eq.(\ref{AA}) we find a lower
limit for the parameter $\beta$ given by
$\beta\gg\frac{\alpha^{3/4}N^{\frac{5C-6\alpha}{4C}}}{(CV_0^2\kappa^{1/3})^{3/4}}$.
In particular for $N=60$, $\alpha=4$ and assuming that the
potential at the end of inflation $V(N=60)\simeq10^{-11}M_p^4$ (in
which $V_0\simeq1.5\times 10^{-12}M_p^4$), we find that the lower
bound for the parameter $\beta$ is given by
$\beta\,M_p^{11/2}\simeq 2.9\times 10^{18}\sim
\,\mathcal{O}(10^{18})$.

Now, combing Eqs.(\ref{NF}), (\ref{pot1}) and (\ref{g1}), we find
that the  relationship  between the e-folding  $N$ and the scalar
field $\phi$ results
\begin{equation}
N=N_0\,(\phi-\phi_0)^{\gamma_1},\label{N1}
\end{equation}
where $\phi_0$ corresponds to an integration constant and the
quantities  $N_0$ and $\gamma_1$ are given by
$$
N_0=\left(\frac{(4C-3\alpha)}{3C}\,\left[\frac{\kappa^2Cg_0V_0}{\alpha}\right]^{1/3}\right)^{\gamma_1},\;\;\;\;\;\mbox{and}
\,\,\,\,\,\gamma_1=\frac{3C}{4C-3\alpha}\,,
$$
respectively.

In this form, we find that the reconstruction of the  effective
potential $V(\phi)$ can be written as
\begin{equation}
V(\phi)=\bar{V}_0\,(\phi-\phi_0)^{\frac{\alpha\,\gamma_1}{C}},\label{VV2}
\end{equation}
where the quantity $\bar{V}_0$ is defined as $\bar{V}_0=V_0\,N_0$.
Here, we note that the effective potential $V(\phi)$ corresponds
to a power law potential in which the exponent
$\frac{\alpha\,\gamma_1}{C}>0$. In particular for case in which
the parameter  $\alpha=4$, we find that the effective potential
$V(\phi)\propto \phi^{0.52}$.

Analogously, from Eqs.(\ref{g1}) and (\ref{N1}), we obtain  that the
coupling parameter $g$ in terms of the scalar field results
\begin{equation}
g(\phi)=\bar{g}_0\,(\phi-\phi_0)^{-\,\gamma_2}=
\bar{g}_0\,\frac{1}{(\phi-\phi_0)^{\gamma_2}},\label{gg2}
\end{equation}
where the constant
$\gamma_2=\frac{6(C-\alpha)}{4C-3\alpha}>0$ and the quantity
$\bar{g}_0=g_0\,N_0$. We noted that in particular if the parameter
$\alpha=4$, then the power $\gamma_2\simeq 1.2$ and the
coupling parameter $g$ decays as $g(\phi)\propto \phi^{-1.2}$. Also, we noted
that for values of $\alpha\simeq 0$ (or equivalently $r\simeq 0$) and the power
$\gamma_2\simeq 3/2$, then   the coupling parameter decays as
 $g(\phi)\propto \phi^{-3/2}$. Also, we note that for the value $\gamma_2\simeq 3/2$
 (or equivalently $\alpha\simeq 0$), the scalar potential $V(\phi)\sim$ constant, leading  to
 an exponential expansion \cite{R1}. Thus, we find that the range for the
 parameter $\gamma_2$ is given by $1.1<\gamma_2 \lesssim3/2$.

In order to give an account of the end of inflationary stage in
this reconstruction, we consider that inflation ends when the
slow-roll parameter $\epsilon_1(\phi=\phi_e)=-\dot{H}/H^2=1$ or
equivalently $\ddot{a}=0$. Under slow roll approximation and
considering the limit ${\cal{A}}\gg 1$, we write $\epsilon_1$ in
terms of standard slow roll parameter $\epsilon$ as
$\epsilon_1\simeq\frac{\epsilon}{(g\,V_\phi)^{1/2}}$.
 In this way, combining Eqs.(\ref{CTM}),  (\ref{VV2}) and (\ref{gg2}), we obtain that the scalar field at the end of inflation becomes
$$
\phi_e=\phi_0+\tilde{\phi_0},\,\,\,\,\,\mbox{where}\,\;\;\;\,\,\tilde{\phi_0}=\left[\frac{1}{2\kappa\,(\bar{g_0}\bar{V_0})^{1/2}}\,
\left(\frac{\alpha\gamma_1}{C}\right)^{3/2}\right]^{\frac{2}{3+\alpha\gamma_1/C-\gamma_2}}>0.
$$
Also, the condition for that inflation takes place is
$\epsilon_1<1$ (or equivalently $\ddot{a}>0$).
 Thus, during inflation the   scalar field is such that
 $\phi>\phi_0+\tilde{\phi_0}$. This  results suggest that
 the coupling parameter $g(\phi)$ given by Eq.(\ref{gg2}) does not
 present a
 singularity at $\phi=\phi_0$, during the inflationary scenario.
 Thus, at the end of inflation
 $g(\phi_e)=\bar{g_0}/\tilde{\phi_0}^{\gamma_2}$ and the scalar
 potential
 $V(\phi_e)=\bar{V_0}\,\tilde{\phi_0}^{\alpha\gamma_1/C}$.

Other type of the attractor for the ratio $r$  studied in the
literature
 is given by $r=1/N(1+\xi N)$, where $\xi$
is a free parameter\cite{Chiba:2015zpa,Herrera:2018cgi}, such that $\xi>-4/315$ in order to obtain
from BICEP2 and Keck Array Collaborations
$r<0.07$, see Ref.\cite{Ob2}. By  considering this attract together with $n_s$ given by Eq.(\ref{e1}), we find
a transcendental equation
 from Eq.(\ref{NF}) to express the number $N$ in terms of the field $\phi$, wherewith  the reconstruction
 does not work.

 In order to find a simple relationship between the number of e-folding $N$ and
 the scalar field, we consider Eq.(\ref{e1}) together with the attractor $r(N)$ given by
\begin{equation}
r=\frac{C}{N(3+\xi N^{1/3})},\label{r3}
\end{equation}
where $\xi$ corresponds to a constant (dimensionless) and it satisfies the lower bound $\xi>-0.24$, such that in particular
 $r(N=60)<0.07$. Note that the relation between the ratio
 $r$ and the scalar spectral index $n_S$, can be written as
\begin{equation}
r(n_s)=\frac{C(1-n_S)^{4/3}}{2[3(1-n_S)^{1/3}+2^{1/3}\xi]}.\label{r4}
\end{equation}
Here, we have used Eqs.(\ref{e1}) and (\ref{r3}), respectively.

 From Eq.(\ref{Vw}) we find that the effective potential $V(N)$
 becomes
 \begin{equation}
 V(N)=\frac{C_1\,N^{1/3}}{3+\xi N^{1/3}},\label{pot2}
\end{equation}
where $C_1>0$ is an integration constant (with units of $M_P^4$). Note that for the case in which the
parameter
 $\xi\gg1 /N^{1/3}$, the scalar potential $V(N)\sim C_1/\xi=$ constant.  By using Eq.(\ref{GGd}) we obtain that
the coupling parameter $g$ in terms of the number of e-folds $N$ results
 \begin{equation}
g(N)=\frac{g_0}{N^{4/3}},\;\;\;\;\mbox{where}\,\,\,\,\,g_0=\beta^2\,C_1^2.\label{g3}
 \end{equation}
Here, we note that the coupling $g(N)$ does not depend of the constant $\xi$,
only of the integration constants $\beta$ and $C_1$, respectively.
From the condition in which  predominate the Galileon effect, where
${\cal{A}}\gg1$, we find a lower bound for the integration constants $\beta $ and $C_1$ given by
 \begin{equation}
g_0\,C_1\sqrt{\kappa}\gg(27N^3+27\xi N^{10/3}+9\xi^2
N^{11/3}+\xi^3 N^{4})^{1/2}.\label{40}
 \end{equation}
In particular for $N=60$ and $\xi=-0.2$ the lower limit gives $g_0C_1\sqrt{\kappa}\gg1534$
and for the value $\xi=0.2$ corresponds to $g_0C_1\sqrt{\kappa}\gg3419$. Also,
in the special case in which the potential at the end of inflation
$V(N=60)=10^{-11}M_p^4$, together with $\xi=-0.2$ we have $C_1\simeq 5\times 10^{-12}M_p^4$,
then we find that the lower limit
for the integration constant $\beta$ is given by $\beta
M_p^{11/2}\gg\,\mathcal{O}(10^{18})$.
For the case $\alpha=0.2$, we have $C_1\simeq 9.7\times10^{-12}M_p^4$ and then  $\beta
M_p^{11/2}\gg 2\times 10^{18}\sim\,\mathcal{O}(10^{18})$.

On the other hand, from Eq.(\ref{NF}) is easy to find that the relation between
$N$ and $\phi$ is given by lineal equation i.e., $N\propto \phi$.
Thus, the reconstruction of  the coupling parameter $g$ as a
function of the scalar field is given by $g(\phi)\propto
\phi^{-4/3}$. This result  suggests that again the parameter
$g(\phi)$ has a behavior power law with  a negative  power. In
this point, we mention that in order to obtain the lineal relation
$N\propto\phi$, we establish from Eq.(\ref{NF}) the condition
$(V_N/(\kappa^2gV^2))^{1/3}=$ constant. Thus, combing
Eqs.(\ref{ns1}) and (\ref{gr8}) together with the attractor $n_S$
given by Eq.(\ref{e1}),  we obtained the tensor to scalar ratio
$r$ given by Eq.(\ref{r3}) (or the potential Eq.(\ref{pot2})).

This methodology can be used for any function $F(N)$, such that
$(V_N/(\kappa^2gV^2))^{1/3}=F(N)$. In this sense, the Eq.(\ref{NF})  takes of
form
  $\int F(N) dN=\int d\phi$, being possible to choose any function $F(N)$ in order to obtain an analytical and invertible solution
  for $N=N(\phi)$. Subsequently, we solve the differential equation for the variable $r(N)$
  (or $V(N)$ or also $g(N)$), by combining Eqs.(\ref{ns1}), (\ref{gr8}) and (\ref{e1})
  for a specific function $F(N)$.
 In particular, for the case in which the function
  $F(N)=$constant, we have $N\propto\phi$, and then we  found that the tensor to scalar
 ratio $r(N)$ is given by Eq.(\ref{r3}), and hence the form of the potential
 $V(N)$ and the coupling $g(N)$ correspond to  the equations
   (\ref{pot2}) and (\ref{g3}), respectively.

\begin{figure}[th]
\includegraphics[width=2.7in,angle=0,clip=true]{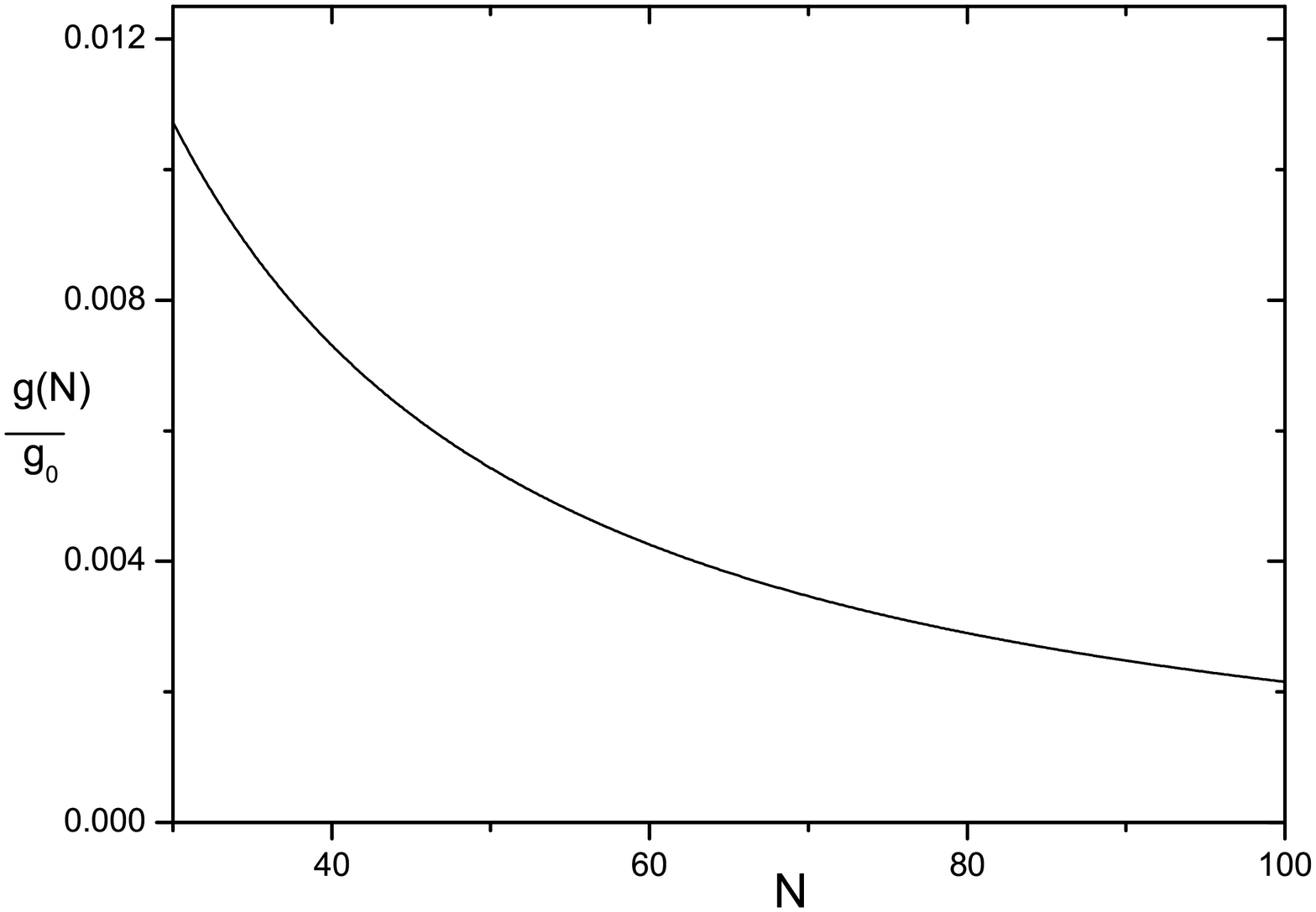}
\includegraphics[width=3.2in,angle=0,clip=true]{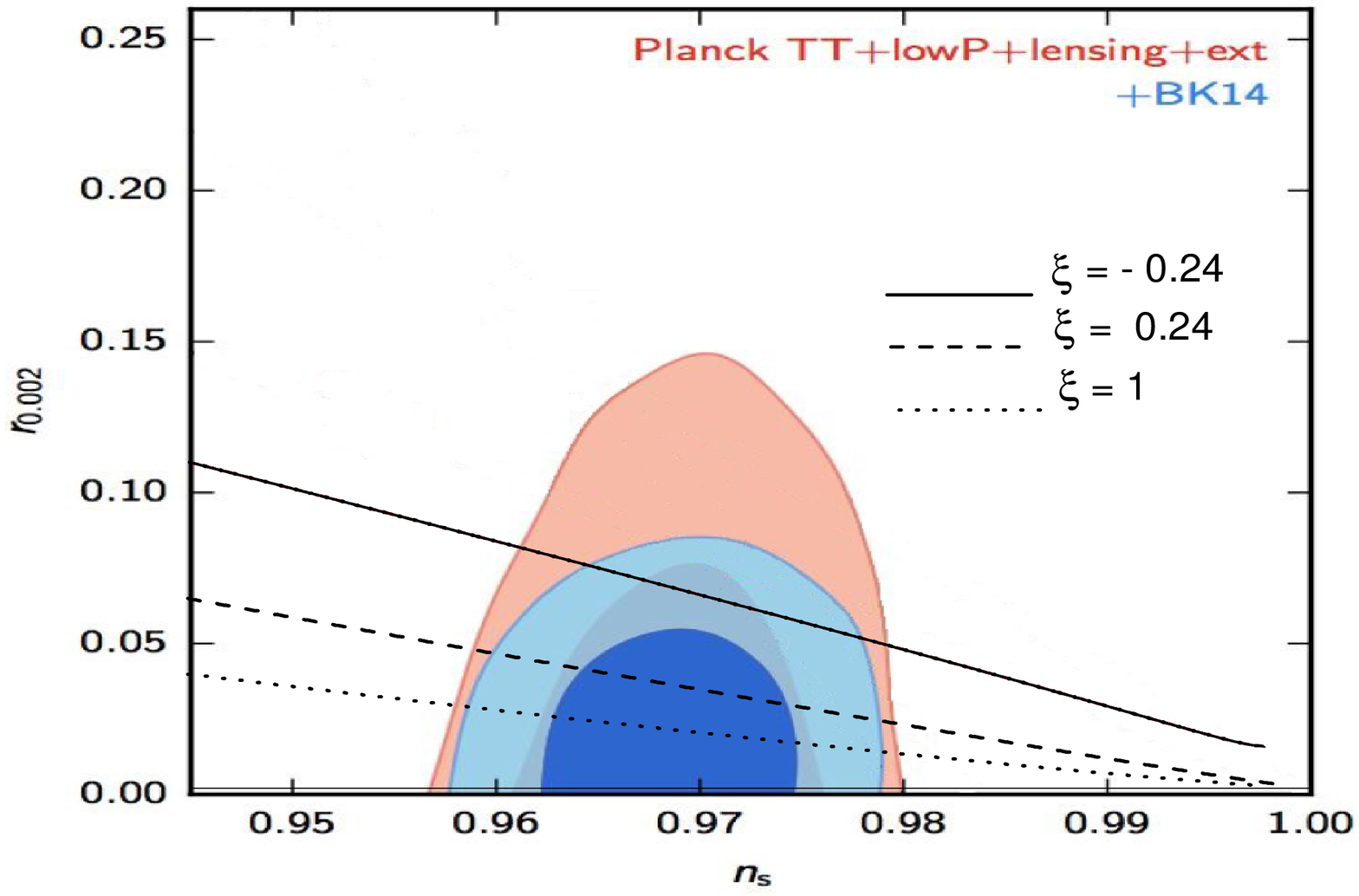}
{\vspace{-0.1 cm}\caption{ The evolution of the ratio $g(N)/g_0$
versus the number of e-folding $N$ (left panel) from
Eq.(\ref{g3}). In the right panel, the plot shows the tensor to
scalar ratio $r$ versus the spectral index $n_S$, from
Eq.(\ref{r4}) for three values of the parameter $\xi$. Here, we
show the two-marginalized constraints on the curve $r=r(n_S)$ and
as the BICEP2 and Keck Array Collaborations data places stronger limits on the  $r-n_S$ plane
\cite{Ob2}.
 \label{fig1}}}
\end{figure}

In Fig.\ref{fig1} we show the evolution of the ratio $g(N)/g_0$ on
the number of e-folds $N$ (left panel) from Eq.(\ref{g3}) and as
it decays in terms of the number of e-folds $N$ (or equivalently
$\phi$, since $N\propto \phi$).
 In the right panel, we show the dependence of the inflationary
parameters $r=r(n_S)$ from Eq.(\ref{r4}). Here, we show the
two-dimensional marginalized constraints (68$\%$ and 95$\%$ C.L.)
on the tensor to scalar ratio $r=r(n_s)$, derived from
observational data\cite{Ob2}. In the plot we have used Eq.(\ref{r4}),
together with three different values of the parameter $\xi$. We
note that for values of $\xi>-0.241$ the relation given by
Eq.(\ref{r4}) is well supported by the BICEP2 and Keck Array Collaborations in which  $r<0.07$
and $n_S\simeq0.967$ \cite{Ob2}.

\section{Conclusions \label{conclu}}

 In this paper we have investigated the reconstruction
 in  the model of G-inflation, from the cosmological parameters such as the scalar
 spectral index
  $n_S(N)$ and the tensor to scalar ratio  $r(N)$, in which $N$ denotes  the
number of e-folding. By assuming the domination of the  Galilean term
 and under the slow roll approximation,
we have developed
 a general formalism of reconstruction in which the function $G(\phi,X)=g(\phi)\,X$ .

 Under
this general analysis we have found  from the attractor point
 $n_S(N)$ and $r(N)$,  integrable expressions
 for the effective potential $V(N)$ and the coupling parameter $g(N)$, respectively.
 Curiously,  in the context of the reconstruction
  we have found that   the tensor to scalar ratio $r$  is similar to  the one
  obtained in the GR, in which $r$ depends only of the potential and its
  derivative, and it does not depend of the coupling $g$.

As the simplest  example and in order to find  the reconstructions
of the  potential $V(\phi)$ and the coupling  parameter $g(\phi)$, we have
assumed  the standard attractors given by $n_S-1=-2/N$ and
$r=\alpha/N$, for large $N$. From these attractors,
 we have applied  our general formalism and also we have found that both, the potential $V(N)$ and the
 coupling parameter $g(N)$ present a power law relation with $N$.  From the
 condition ${\cal{A}}\gg1$ in which  predominate the Galileon term, we have obtained
  a lower bound for the integration constant
  $\beta$ given by $\beta\gg\frac{\alpha^{3/4}N^{\frac{5C-6\alpha}{4C}}}{(CV_0^2\kappa^{1/3})^{3/4}}$.
  In particular for the specific cases  in which  $N=60$, $\alpha=4$ and
   $V_0/M_p^4\sim \mathcal{O}(10^{-12})$, we have obtained
that the lower limit for the  constant $\beta$  results
  $\beta M_p^{11/2} \sim \mathcal{O}(10^{18})$. In this context, from the
  standard attractor point, we have found that the reconstruction of the scalar
  potential and the coupling parameter are given by Eqs.(\ref{VV2}) and (\ref{gg2}).
  Here, we have determined that the parameter $g(\phi)$ decays as $g(\phi)\propto \phi^{-\gamma_2}$
  with $\gamma_2$ a positive constant.

 Other, important attract in GR corresponds to $r=(N(1+\xi N))^{-1}$ together $n_s-1=-2/N$, here
 we could not find  an analytical expression for $N=N(\phi)$,
  in order to obtain the reconstruction of $V(\phi)$ and
 $g(\phi)$, respectively.

Another, analytical reconstruction in the model of  G-inflation
corresponds to the tensor to scalar ratio $r(N)$ given by
Eq.(\ref{r3}).  Assuming that the function $F(N)=$ constant, we
have find a simple relation between the number $N$ and $\phi$
given by  $N\propto \phi$ and  then  Eq.(\ref{r3}). Here, we have
obtained that the parameter $g$ in terms of the scalar field is
given by $g(\phi)\propto \phi^{-4/3}$ and the effective potential
$V(\phi)\propto \phi^{1/3}/(3+\xi'\phi^{1/3})$.
 Also, in particular we have find that for
the case in which the parameter  $\xi\gg 1/N^{1/3}$,
 the scalar potential $V(\phi)$ becomes  a constant\cite{R1}.
 In fact, by assuming the condition ${\cal{A}}(N)\gg 1$, we have
 obtained a lower bound for the integration constants given by
 Eq.(\ref{40}).


Finally in this paper, we have not addressed the reconstruction to
another  functions $G(\phi,X)$ in the action, such as
$G(\phi,X)=g(\phi)\,X^{n}$ or $g_1(\phi)X+g_2(\phi)X^2+..$ or
other. In this sense,
 we hope
to return to this point in the near future.

\begin{acknowledgments}
The author thanks Prof. Marco Olivares for useful discussions on
the $r-n_S$ plane.
 This work was supported by
Proyecto VRIEA-PUCV N$_{0}$ 039.309/2018.
\end{acknowledgments}



\begin{thebibliography}{99}                                                                                               %

\bibitem {R102}A.A. Starobinsky, Phys. Lett. B \textbf{91}, 99 (1980).
\bibitem {R1}A. Guth , Phys. Rev. D \textbf{23}, 347 (1981).



\bibitem {R103}A.D. Linde, Phys. Lett. B \textbf{108}, 389 (1982);
A.D. Linde, Phys. Lett. B \textbf{129}, 177 (1983).

\bibitem{Rin} K.~Sato,
  Mon.\ Not.\ Roy.\ Astron.\ Soc.\  {\bf 195}, 467 (1981).

\bibitem {R2}V.F. Mukhanov and G.V. Chibisov , JETP Letters \textbf{33},
532 (1981);
S. W. Hawking,Phys. Lett. B \textbf{115}, 295 (1982).

\bibitem {R203}A. Guth and S.-Y. Pi, Phys. Rev. Lett. \textbf{49}, 1110 (1982);
A. A. Starobinsky, Phys. Lett. B \textbf{117}, 175 (1982).



\bibitem {astro}D.~Larson \textit{et al.},
Astrophys.\ J.\ Suppl.\ \textbf{192}, 16 (2011);
C.~L.~Bennett \textit{et al.},
Astrophys.\ J.\ Suppl.\ \textbf{192}, 17 (2011);
N.~Jarosik \textit{et al.},
Astrophys.\ J.\ Suppl.\ \textbf{192}, 14 (2011).

\bibitem{Hinshaw:2012aka}
  G.~Hinshaw {\it et al.}  [WMAP Collaboration],
  Astrophys.\ J.\ Suppl.\  {\bf 208}, 19 (2013).

\bibitem{Ade:2013zuv}
  P.~A.~R.~Ade {\it et al.}  [Planck Collaboration],
  Astron.\ Astrophys.\  {\bf 571}, A16 (2014);
  P.~A.~R.~Ade {\it et al.}  [Planck Collaboration],
  Astron.\ Astrophys.\  {\bf 571}, A22 (2014).

\bibitem{Planck2015}
  P.~A.~R.~Ade {\it et al.} [Planck Collaboration],
  Astron.\ Astrophys.\  {\bf 594}, A20 (2016).

  \bibitem{Ob2} P. A. R. Ade {\it et. al.} [BICEP2 and Keck Array Collaborations], Phys. Rev. Lett.
 {\bf 116}, 031302 (2016).

\bibitem{DiValentino:2016foa}
  E.~Di Valentino {\it et al.} [CORE Collaboration],
  JCAP {\bf 1804}, 017 (2018); F.~Finelli {\it et al.} [CORE Collaboration],
  JCAP {\bf 1804}, 016 (2018).



















\bibitem{Ho}  G. W. Horndeski, Int. J. Theor. Phys. {\bf 10}, 363 (1974).

\bibitem{Kobayashi:2011nu}
  T.~Kobayashi, M.~Yamaguchi and J.~Yokoyama,
  Prog.\ Theor.\ Phys.\  {\bf 126}, 511 (2011).


\bibitem{G1} C. Deffayet, O. Pujolas, I. Sawicki and A. Vikman,
 JCAP {\bf 1010}, 026 (2010).

 \bibitem{G2} T. Kobayashi, M. Yamaguchi and J. Yokoyama,
Phys. Rev. Lett. {\bf 105}, 231302 (2010).





\bibitem{TheLIGOScientific:2017qsa}
  B.~P.~Abbott {\it et al.} [LIGO Scientific and Virgo Collaborations],
  Phys.\ Rev.\ Lett.\  {\bf 119}, no. 16, 161101 (2017).



\bibitem{Monitor:2017mdv} B.~P.~Abbott {\it et al.}
[LIGO Scientific and VIRGO Collaborations],
  Living Rev.\ Rel.\  {\bf 19}, 1 (2016);
  B.~P.~Abbott {\it et al.} [LIGO Scientific and Virgo and Fermi-GBM and INTEGRAL Collaborations],
  Astrophys.\ J.\  {\bf 848}, no. 2, L13 (2017).

  \bibitem{GBM:2017lvd}
  B.~P.~Abbott {\it et al.},
  Astrophys.\ J.\  {\bf 848}, no. 2, L12 (2017).



















\bibitem{Nic}
A. Nicolis, R. Rattazzi, E. Trincherini, Phys. Rev. D {\bf 79}, 064036 (2009).





\bibitem{288}K. Kamada, T. Kobayashi, M. Yamaguchi and
J. Yokoyama, Phys. Rev. D {\bf 83}, 083515 (2011).



\bibitem{289a}K.~Kamada, T.~Kobayashi, T.~Kunimitsu, M.~Yamaguchi and J.~Yokoyama,
  Phys.\ Rev.\ D {\bf 88}, no. 12, 123518 (2013).

  \bibitem{289}K. Kamada, T. Kobayashi, T. Takahashi,
M. Yamaguchi and J. Yokoyama, Phys. Rev. D {\bf 86}, 023504 (2012).



\bibitem{Ut} S.~Hirano, T.~Kobayashi and S.~Yokoyama,
  Phys.\ Rev.\ D {\bf 94}, no. 10, 103515 (2016).


\bibitem{Pw} S.~Unnikrishnan and S.~Shankaranarayanan,
  JCAP {\bf 1407}, 003 (2014).
\bibitem{Herrera:2017qux}
  R.~Herrera,
  JCAP {\bf 1705}, no. 05, 029 (2017).


\bibitem{Br} H.~Bazrafshan Moghaddam, R.~Brandenberger and J.~Yokoyama,
  Phys.\ Rev.\ D {\bf 95}, no. 6, 063529 (2017).




\bibitem{agr} M.~He, J.~Liu, S.~Lu, S.~Zhou, Y.~F.~Cai, Y.~Wang and R.~Brandenberger,
  JCAP {\bf 1612}, no. 12, 040 (2016).

\bibitem{Ka}
Y. F. Cai, J. O. Gong, S. Pi, E. N. Saridakis and S. Y. Wu, Nucl. Phys. B {900}, 517 (2015)

\bibitem{Reh2}J.~Ohashi and S.~Tsujikawa,
  JCAP {\bf 1210}, 035 (2012).







\bibitem{DE1} T.~Kobayashi,
  Phys.\ Rev.\ D {\bf 81}, 103533 (2010).

\bibitem{DE2}
   S. Nesseris, A. De Felice, and S. Tsujikawa, Phys. Rev. D 82, 124054 (2010); A.~De Felice and S.~Tsujikawa,
  JCAP {\bf 1202}, 007 (2012);   D.~Maity and P.~Saha,
  arXiv:1801.08080 [hep-ph]; R.~Herrera, N.~Videla and M.~Olivares,
  arXiv:1806.04232 [gr-qc].


\bibitem{H1}F. Lucchin, S. Matarrese, Phys. Rev. D {\bf 32}, 1316 (1985); R. Easther, Class. Quantum Grav. {\bf 13}, 1775 (1996).

\bibitem{H2}J. Martin, D. Schwarz, Phys. Lett. B {\bf 500}, 1-7 (2001).

\bibitem{H3} X.~z.~Li and X.~h.~Zhai,
  Phys.\ Rev.\ D {\bf 67}, 067501 (2003).

  \bibitem{H4}
  R.~Herrera and R.~G.~Perez,
  Phys.\ Rev.\ D {\bf 93}, no. 6, 063516 (2016).

\bibitem{M}  V.~Mukhanov,
  Eur.\ Phys.\ J.\ C {\bf 73}, 2486 (2013).

\bibitem{Chiba:2015zpa}
  T.~Chiba,
  PTEP {\bf 2015}, no. 7, 073E02 (2015).

\bibitem{H5}T.~Miranda, J.~C.~Fabris and O.~F.~Piattella,
  JCAP {\bf 1709}, no. 09, 041 (2017); A.~Ach\'ucarro, R.~Kallosh, A.~Linde, D.~G.~Wang and Y.~Welling,
  JCAP {\bf 1804}, no. 04, 028 (2018);S.~D.~Odintsov and V.~K.~Oikonomou,
  Annals Phys.\  {\bf 388}, 267 (2018);P.~Christodoulidis, D.~Roest and E.~I.~Sfakianakis,
  arXiv:1803.09841 [hep-th];  P.~Carrilho, D.~Mulryne, J.~Ronayne and T.~Tenkanen,
  arXiv:1804.10489 [astro-ph.CO]; S.~D.~Odintsov and V.~K.~Oikonomou,
  Nucl.\ Phys.\ B {\bf 929}, 79 (2018);  S.~D.~Odintsov and V.~K.~Oikonomou,
  Phys.\ Rev.\ D {\bf 97}, no.6,  064005 (2018).


\bibitem{T}R. Kallosh and A. Linde, JCAP 1307, 002 (2013).

\bibitem{E}R. Kallosh and A. Linde, JCAP 1310, 033 (2013).





\bibitem{Higgs} D. Kaiser,   Phys. Rev. D
{\bf52},  4295ÃÂÃÂ4306 (1995); F Bezrukov  and M.
Shaposhnikov, Phys Lett B {\bf659},  703, (2008).


\bibitem{Higgs2} R. Kallosh, A. Linde, D. Roest,   Phys Rev
Lett, {\bf 112} 011303, (2014).



\bibitem{Herrera:2018cgi}
  R.~Herrera,
  Eur.\ Phys.\ J.\ C {\bf 78}, no. 3, 245 (2018).


\bibitem{Huang:2007qz}
  Q.~G.~Huang,
  Phys.\ Rev.\ D {\bf 76}, 061303 (2007).




\bibitem{Gao:2017owg} J.~Lin, Q.~Gao and Y.~Gong,
  Mon.\ Not.\ Roy.\ Astron.\ Soc.\  {\bf 459}, no. 4, 4029 (2016);
  Q.~Gao,
  Sci.\ China Phys.\ Mech.\ Astron.\  {\bf 60}, no. 9, 090411
  (2017).


\bibitem{Roest:2013fha}
  D.~Roest,
  JCAP {\bf 1401}, 007 (2014).


\bibitem{N1}J.~Garcia-Bellido and D.~Roest,
  Phys.\ Rev.\ D {\bf 89}, no. 10, 103527 (2014).

\bibitem{N2}P.~Creminelli {\it et al.},
  Phys.\ Rev.\ D {\bf 92}, no. 12, 123528 (2015).


\bibitem{GG1}X. Gao and D. A. Steer, JCAP 1112, 019 (2011).

\bibitem{GG2}A. De Felice and S. Tsujikawa, Phys. Rev. D 84, 083504 (2011).

\bibitem{Ohashi:2012wf}
  J.~Ohashi and S.~Tsujikawa,
  JCAP {\bf 1210}, 035 (2012).
























































































































































  \bibitem{go}A.~De Felice, T.~Kobayashi and S.~Tsujikawa,
  Phys.\ Lett.\ B {\bf 706}, 123 (2011); C.~Burrage, C.~de Rham, D.~Seery and A.~J.~Tolley,
  JCAP {\bf 1101}, 014 (2011).




 %






























\end{thebibliography}
\end{document}